# Human Reliability Analysis for Oil and Gas Operations: Analysis of Existing Methods

**Marilia A. Ramos**
University of California Los Angeles (UCLA)
Engineering VI, 404 Westwood Plaza, Los Angeles -CA, 90095
Marilia.ramos@ucla.edu

**Camille Major**
Chevron Technology, Projects and Services
1400 Smith Street, Houston – TX, 77002
CMajor@chevron.com

**Nsimah Ekanem**
Chevron Technology, Projects and Services
1400 Smith Street, Houston – TX, 77002
Nsimah.Ekanem@chevron.com

**Cesar Malpica**
Chevron Technology, Projects and Services
1400 Smith Street, Houston – TX, 77002
Cesar.Malpica@chevron.com

**Ali Mosleh**
University of California Los Angeles (UCLA
Enginnering VI, 404 Westwood Plaza, Los Angeles -CA, 90095
mosleh@ucla.edu

## Abstract

In the petroleum industry, Quantitative Risk Analysis (QRA) has been one of the main tools for risk management. To date, QRA has mostly focused on technical barriers, despite many accidents having human failure as a primary cause or a contributing factor. Human Reliability Analysis (HRA) allows for the assessment of the human contribution to risk to be assessed both qualitatively and quantitatively. Most credible and highly advanced HRA methods have largely been developed and applied in support of nuclear power plants control room operations and in context of probabilistic risk analysis. Moreover, many of the HRA methods have issues that have led to inconsistencies, insufficient traceability and reproducibility in both the qualitative and quantitative phases. Given the need to assess human error in the context of the oil industry, it is necessary to evaluate available HRA methodologies and assess its applicability to petroleum operations. Furthermore, it is fundamental to assess these methods against good practices of HRA and the requirements for advanced HRA methods. The present paper accomplishes this by analyzing seven HRA methods. The evaluation of the methods was performed in three stages. The first stage consisted of an evaluation of the degree of adaptability of the method for the Oil and Gas industry. In the second stage the methods were evaluated against desirable items in an HRA method. The higher-ranked methods were evaluated, in the third stage, against requirements for advanced HRA methods. In addition to the methods' evaluation, this paper presents an overview of state-of-the-art discussions on HRA, led by the Nuclear industry community. It remarks that these discussions must be seriously considered in defining a technical roadmap to a credible HRA method for the Oil and Gas industry.

# 1 Introduction

Petroleum installations pose safety concerns that are inherent to their characteristics—after all, operations and day-to-day work happen among flammable and toxic fluids. Although the oil and gas industry has made advances to improve safety, accidents of all ranges still occur. Since 1992, when the Process Safety Management standard (PSM) was promulgated by the Occupational Safety and Health Administration (OSHA), no other industry sector has had as many fatal or catastrophic incidents related to the release of highly hazardous chemicals (HHC) as has been the case with the petroleum refining industry [1]. Statistics show that the majority of accidents (over 80%) in the chemical and petrochemical industries are related to human failure [2]. The American Petroleum Institute states that in the recent decades, the 100 largest accidents at chemical and hydrocarbon processing facilities have severely injured or killed hundreds of people, contaminated the environment, and caused several property damage. Moreover, human error was a significant factor in almost all these accidents. In systems where a high degree of hardware redundancy minimizes the consequences of single component failures, human errors may comprise over 90 percent of the system failure probability [3].

In the petroleum industry, Quantitative Risk Analysis (QRA) has been one of the main tools for risk management. QRAs differ on the extent to which they incorporate human and organizational factors. To date, QRA has mostly focused on technical barriers [4–6]. It is important to note that the need for analyzing human error in the process industry is recognized by the American Petroleum Institute, in the guide "A Manager's Guide to Reducing Human Errors (API 770)" [3]. The guide states: "any serious attempt to improve process safety must address the fact that human errors in the design, construction, operation, maintenance, and management of facilities are the root causes of almost all quality deficiencies, production losses, and accidents".

Human Reliability Analysis (HRA) allows for identifying, modeling and quantifying human errors, their causes, and their consequences. The results of an HRA can be used for developing risk reduction measures and make risk-informed decisions. HRA has deep roots in the Nuclear industry, and many HRA methods that exist nowadays were developed for and/or primarily applied to Nuclear Power Plants (NPPs) operations. Oil and Gas operations and NPPs have, however, remarkable differences concerning operation, control room layouts, time window for accident avoidance and mitigation, training and safety culture [7]. Taylor et al. [5] provide an overview of these differences. In the lack of a method specific for oil and gas operations, the industry has been generally applying generic HRA methods, which may not correctly represent the peculiarities of their operations.

Recent initiatives aim at developing an HRA method tailored for the Oil and Gas industry, such as the Petro-HRA project [8] and Phoenix for Petroleum Refining Operations (Phoenix-PRO) [7]. Both methods consist of adaptations of existing HRA methods. Petro-HRA method modified the SPAR-H method [9] for representing oil and gas operations, with a focus on upstream operations. Phoenix-PRO modified Phoenix HRA method [10] for oil refineries and petrochemical plants operations.

The features of an HRA method for the Oil and Gas industry should not stop at its suitability for representing and analyzing its operations. It is equally important that the model is robust qualitatively and quantitatively; that it provides complete guidance for application, and that it generates transparent, traceable and reproducible results. Many methods have been largely reviewed and discussed for their advantages and limitations [11] or assessed against operational experience benchmarking studies [12]. Conclusions of these studies have identified that several methods have limitations regarding qualitative and quantitative aspects. The experience of the Nuclear industry, primary user of HRA, should be leveraged by the Oil and Gas industry regarding desirable features on an HRA method. The state of the art regarding HRA and the requirements for a robust method must be seriously considered in defining a technical roadmap to a credible HRA method for the industry.

Given the need to assess human error in the context of oil and gas operations, it is necessary to evaluate existing HRA methodologies and i) assess their applicability to oil and gas operations; ii) evaluate their features against HRA good practices and requirements for a robust HRA method. An HRA methodology designed specifically for the context of petroleum industry operations would allow for the identification of the peculiarities of this industry regarding interactions between the crew and the plant, possible operators' errors and contextual factors. This would be possible given the combination of the benefits brought by a robust HRA model considering the inherent characteristics of the petroleum sector in terms of the type of work, the individuals and the organization that influence human performance, operators training and organization, and operating procedures.

This paper aims at evaluating existing HRA methods for their applicability in oil and gas, to assess their potential to serve as a foundation for a method for this industry. The evaluation is carried out in three stages. First, the methods were evaluated concerning the efforts related to adaptations for Oil and Gas operations. Second, the methods were evaluated regarding desirable features of an HRA method, concerning HRA good practices. Third, high-ranked methods were evaluated against state-of-the-art requirements for advanced HRA.

This paper is organized as follows. Section 2 provides an overview on the concepts and history of HRA. Section 3 elaborates on good practices of HRA application and the requirements for advanced HRA methods. Section 4 provides an overview of the HRA methods selected for the evaluation. These methods are evaluated in Section 5. Finally, Section 6 summarizes the discussion and presents concluding thoughts.

## 2    Human Reliability Analysis: Overview

Human reliability can be defined as the probability that a person (1) performs an action correctly as required by the system in a required time and (2) that this person does not perform any extraneous activity that can degrade the system [13]. Studies on human reliability are a relatively recent area of research; its history can be traced back to 1952, when it was first addressed for a weapon system feasibility in Sandia National Laboratories, USA [14]. In the 1960s, in turn, it started being used for civil applications, with a focus on man machine system design.

The first formal method for HRA was presented in November 1962 at the Sixth Annual Meeting of the Human Factors Society, followed by a monograph from Sandia Laboratories [15] outlining its quantification. This method, called Technique of Human Error Rate Prediction (THERP), is still used in HRA [16]. Throughout the 1980s the number of HRA methods significantly increased. It is possible to observe a strong correlation between the accident at Three-Mile Island (TMI)$^{(*)}$ on March 28, 1979—which was the most serious accident in the U.S. commercial nuclear plant operating history—and the growth in the number of HRA methods [17]. Figure 1 below presents the evolution of the number of HRA methods through time. Although there are currently many HRA methods, some of them are variations of the same approach, and the number of significantly different HRA approaches is therefore smaller than it seems.

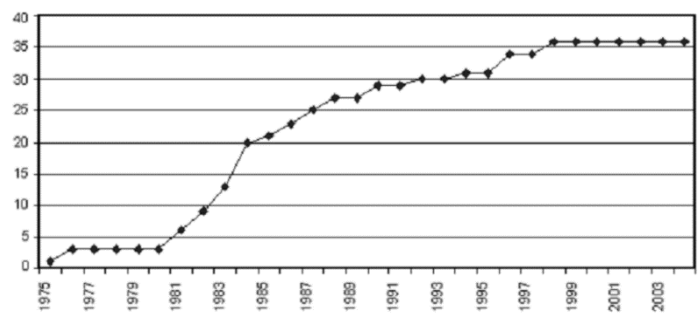

Figure 1: Cumulated number of HRA methods according to year of publication [17]

It is important to notice that although human error is the main object of study in HRA, it must be analyzed considering its possible causes and context. Indeed, one of the undisputed assumptions in HRA approaches is that the quality of human performance depends on the conditions under which the tasks or activities are carried out [18]. These conditions, in turn, have generally been referred to as Performance Shaping Factors (PSFs) or Performance Influencing Factors (PIFs); and they serve to either enhance or degrade human performance relative to the expected action. PIFs can reflect organizational matters – e.g. as safety culture; personal – e.g. motivation, and others. Human error can be generally categorized as errors of omission—when an operator fails in acting, i.e., she/he does not perform a specific task, and errors of commission—when an operator fails when acting, i.e., she/he performs a task in a wrong manner, or perform the wrong task, or in the wrong timing.

---

$^{*}$ The Three-Mile Island accident was the most serious accident in U.S. commercial nuclear power plant operating history. The TMI-2 reactor partially melted down. Although this is the most dangerous kind of nuclear power accident, its small radioactive releases had no detectable health effects on plant workers or the public

Despite the several differences among HRA methods, HRA is often depicted as consisting of three distinct phases [19]:

> 1. Modeling of the potential human errors—the enlistment of some variety of task analysis to decompose an overall sequence of events into smaller units suitable for analysis. There is no universally agreed standard for the best level of decomposition.
>
> 2. Identification of the potential contributors to human error—the selection of relevant performance shaping factors. As with task decomposition, there is no standard list of performance shaping factors, and there is considerable variability between HRA methods.
>
> 3. Quantification of human errors—the calculation of a human error probability (HEP). Each HRA method features a different approach to quantification, including expert estimation, the use of PSF multipliers, Bayesian approaches, and simulations. The quantification determines the likelihood that the particular action modeled in the previous steps will fail.

Most HRA specialists classify HRA methods as first and second generation. Although there is no consensus in the literature when it comes to such classification, the HRA community has been generally more inclined to make use of the generational classification and has done so simply in terms of chronology. The oldest, first developed HRA methods are colloquially considered first generation methods, while subsequent methods—the descendants of the earlier methods—are considered second-generation methods.

In short, HRA makes it possible for the human contribution to risk to be assessed both qualitatively and quantitatively. Its use allows to identify, model and quantify human failure events (HFE) in the context of various accident scenarios. Such analyses can, in turn, form the basis for prioritizing and developing effective safeguards to prevent or reduce the likelihood of human caused accidents.

## 3  Human Reliability Analysis: General Requirements

HRA has evolved as a discipline throughout the past four decades. Yet, many HRA methods still present shortcomings among which are lack of traceability and reproducibility of the analysis. The current efforts by the HRA community to address these shortcomings benefit from a *momentum* that combines the following:

1. Increasing understanding of human behavior. Second generation HRA advanced the analysis of human error by including aspects of human cognitive response into the methods. Since then, there is a growing understanding that HRA should benefit from the areas of cognitive and behavioral sciences. The inclusion of these sciences improves HRA's ability to correctly model how operators interact with the system, how they could fail, and for which reasons.
2. HRA data collection initiatives. Recent efforts in data collection for use in HRA include the development of the SACADA database[†]. In addition to expert judgement and cognitive sciences, HRA methods can now use extended empirical data for modeling and quantifying human actions.
3. Development of more robust quantification methods. Some HRA methods use simple approaches for quantification, such as a multiplicative method, adjusting a baseline number to account for context factors. Although these may be useful in some applications, HRA can benefit from more robust quantification frameworks. Methods such as Bayesian Belief Networks (BBNs) are increasingly used in HRA [20]. The increasing adoption of robust quantification methods is highly linked to the $4^{th}$ point:
4. Access and improvement of computational power. Easy access to powerful software and hardware allows for performing HRA using more complex models without necessarily increasing the time for analysis.
5. Accumulated experience of HRA use. The Nuclear industry have been developing and using a variety of HRA methods for a long time. The industry has now an extensive accumulated experience on the advantages and limitations of the methods and their application.

---

[†] The Scenario Authoring, Characterization, and Debriefing Application (SACADA) database was developed by the U.S. NRA to address the data need in HRA. More information on SACADA can be seen in [48].

The points above are a common topic in advancing the state of the art among the Nuclear industry HRA community and can be leveraged by the oil and gas industry when developing / adapting HRA methods. The oil and gas industry may not share exactly the same requirements as the Nuclear industry (for example on data collection) at the moment, but almost all the points above need to be seriously considered in defining a technical roadmap to a credible HRA method for the industry. Clearly even in areas where the activity (data gathering) may seem to be specific to nuclear application, there are many fundamentals that transcend domain-specific applications (for example, data indicating influences of performance influencing factors of cognition and decision making in a Control Room setting).

In addition to the above features that can be leveraged by the Oil and Gas Industry, this section presents an overview of good practices for implementing HRA and requirements for advanced HRA methods. The aim is to provide a foundation for the evaluation of the HRA methods in Section 5. Three approaches are presented. Good practices for implementing HRA, set by the United States Nuclear Regulatory Commission of the (U.S. NRC), in Section 3.1, followed by the need and requirements of a model-based HRA method [21] in Section 3.2, and the requirements for third generation HRA [22] in Section 3.3. These approaches are interconnected and point to the same direction. The two last approaches were chosen as they combine state of the art discussions from a diversity of publications to draw requirements for advanced credible HRA methods. Section 5 will present an evaluation of selected HRA methods with respect to some key features drawn by these approaches.

*3.1 Good Practices for Implementing HRA – NUREG-1792.*

The Good Practices for Implementing HRA (NUREG-1792) [23] was developed by the Sandia National Laboratories and the Nuclear Regulatory Commission of the United States (U.S. NRC). Although the development of this report was in the context of Probabilistic Risk Assessment (PRA) for commercial NPP operation, it contains valuable discussions and conclusions on general features of the methods. The report states that laying good practices for implementing HRA was driven by lack of consistency among HRA practitioners on the treatment of human performance in the context of a PRA. The report also adds that, depending on the application, a given analysis may not have to meet every good practice. With the good practices in mind, practitioners can determine the extent to which an analysis is adequate. An evaluation of the HRA methods under the light of this report is further developed in "Evaluation of Human Reliability Analysis Methods Against Good Practices" – NUREG-1842 [24].

NUREG-1792 identifies 38 Good Practices for HRA. Some of these Good Practices are summarized below:

1. The HRA assessment should involve a multi-disciplinary team and should include field observations, review of plant documents and talk-throughs with the plant crew;
2. The analysis should account for dependencies among the Human Error Probabilities in an accident sequence;
3. HRA should address both diagnosis and response execution failures;
4. Plant- and Activity-Specific PSFs should be included in the Detailed Assessments;
5. The analysis should define and consider recovery actions;
6. The analysis should address Errors of Commission in addition to Errors of omission;
7. HRA application should be documented well enough so that its results are traceable and reproducible.

*3.2 The need of a model based HRA method*

Driven by the limitations of HRA methods, Mosleh & Chang [21] summarize the need and the basis for developing requirements for the next generation HRA methods. They argue that a model-based approach is an answer for the inconsistencies, insufficient traceability and reproducibility in both the qualitative and quantitative phases of HRA methods and should thus be at the core of the advanced methods.

To overcome the general shortcomings of HRA methods, and reflecting expectations from various authors, Mosleh and Chang [21] state that an HRA method should enable analysts to:

- identify human response (errors are the main focus);
- estimate response probabilities; and,
- identify causes of errors to support development of preventive or mitigating measures.

Moreover, the method should:

- include a systematic procedure for generating reproducible qualitative and quantitative results,
- have a causal model of human response with roots in cognitive and behavioral sciences, and with elements (e.g. PIFs) that are directly or indirectly observable, and a structure that provides unambiguous and traceable links between its input and output
- be detailed enough to support data collection, experimental validation, and various applications of PSA[‡]. Data and model are two tightly coupled entities.

In addition, they add that that the model should be data-informed, and conversely the data collection and analysis must be model-informed. In that direction, they note that:

---

[‡] A parallel can be ellaborated for the Oil and Gas indutry: the method should support different applications of QRA.

- Only a causal model can provide both the explanatory and predictive capabilities. Without a causal model it is difficult for instance to explain why in some cases seemingly similar contexts result in different outcomes;
- Only a model-based approach provides the proper framework for tapping into and integrating models and data from the diverse scientific disciplines that address different aspects of human behavior;
- A causal model that explicitly captures the generic and more fundamental aspects of human response can be tested and enhanced using data and observations from diverse context. This is particularly important as the situations of interest in HRA are highly context-dependent and rare, meaning that adequate statistical data are unlikely to be available for a direct estimation of operator response probabilities;
- A generic causal model will have a much boarder domain of applicability, reducing the need for developing application-specific methods. For instance, the same underlying model can be used for errors during routine maintenance work, as well as operator response to accidents;
- Only a model-based method can ensure reproducibility of the results, and robustness of the predictions.

Considering the above, efforts were taken into developing a model-based HRA method [25–28], which later became the foundation of Phoenix HRA method [29,30].

### *3.3 Developing Third Generation HRA Methods*

In the past 10 years, two directions characterize most of the efforts by the HRA community to modernize HRA: data collection and the development of HRA methods based on cognitive science. Groth et al. [22] remark that these initiatives have largely proceeded independently, and there has been little research into how to leverage advances in data gathering with the scientific advances in modeling and analysis methods. To bridge this gap, they summarize requirements of 3$^{rd}$ Generation HRA methods, which should combine simulator data$^§$, causal models and cognitive science. These requirements are a result of a combination of many HRA studies and previous research into use of HRA data.

According to their assessment, a third generation HRA method must be:

- Comprehensive
    - C1: Clearly acknowledge that HRA addresses a joint system of humans and machines addressing the needs of an engineering system. In that sense,
        - a human failure event (HFE) is an event reflecting a process, not a single action;
        - an HFE is a failure of a team working in an organizational context, not of a single human;
        - an HFE can be caused not only by human errors, but also by machine failures which can deprive the humans of required information or control abilities.

---

$^§$ Simulator data for HRA can be obtained through the use of simulators of the plant and analysis of how operators react in hazardous situations, when interacting with the simulator.

- - o C2: Provide explicit representation of the causal factors that affect human–machine team failures based on both cognitive science and systems engineering.
    - o C3: Address the full spectrum of contexts and causal factors that are relevant to HRA.
  - Research-based
    - o RB1: Be informed by both data and models, and spanning multiple scientific and engineering disciplines, and multiple scales.
    - o RB2: Be able to incorporate multiple sources, types, and sizes of data, models, and information.
  - Adaptable and flexible
    - o AF1: Be designed to be updated at reasonable engineering intervals.
    - o AF2: Be flexible enough to accommodate changes in database structures, data sources, and methodologies.
  - Multi-Purpose
    - o MP1: Support the quantitative and qualitative aspects of HRA as a part of probabilistic risk assessment (PRA), including quantification of Human Error Probabilities (HEPs).
    - o MP2: Provide quantitative insights that go beyond HEP quantification, to enable improving human–machine team performance.

# 4 Human Reliability Analysis Methods Overview

This section presents a brief review of existing HRA methods. Given the large number of existing methods, this review considered the following criteria for methods' selection:

- The method must be publicly available in terms of documentation and publications;
- The method must have recent documentation;
- The method must have indication of recent use or development in the literature, and
- The method must provide rules for calculating human error probabilities (HEPs).

The criteria abovementioned allowed for the selection of the most promising methods for the purpose of this study. This review thus encompasses the following methods:

1. THERP (Technique for Human Error Rate Prediction)
2. CREAM (Cognitive Reliability and Error Analysis Method)
3. HEART (Human Error Assessment and Reduction Technique)
4. ATHEANA (A Technique for Human Event Analysis)
5. SPAR-H (Standardized Plant Analysis Risk HRA method)
6. PETRO-HRA
7. PHOENIX-PRO

Despite following the criteria of the study, two methods were not included in the review: ASEP, for being an abbreviated and slightly modified version of THERP [31], and SLIM-MAUD [32], for being a quantitative method only.

The overview leverages from published reports and recognized papers on the subject, in addition to the methods' guidelines and handbooks. This report does not intend to detail the methods' procedures. For a deeper understanding on how to apply the methods, the reader can refer to the methods' manuals.

*4.1 THERP (Technique for Human Error Rate Prediction)*

THERP is known as the first formal HRA method. Its handbook was prepared by Swain and Guttman in 1983 [13] for the US Nuclear Regulatory Commission (NRC). The aim of THERP is to calculate the probability of successful performance of activities needed for the execution of a task. The results are represented graphically in an HRA event tree, which is a formal representation of required actions sequence needed, as can be seen in Figure 2 [33].

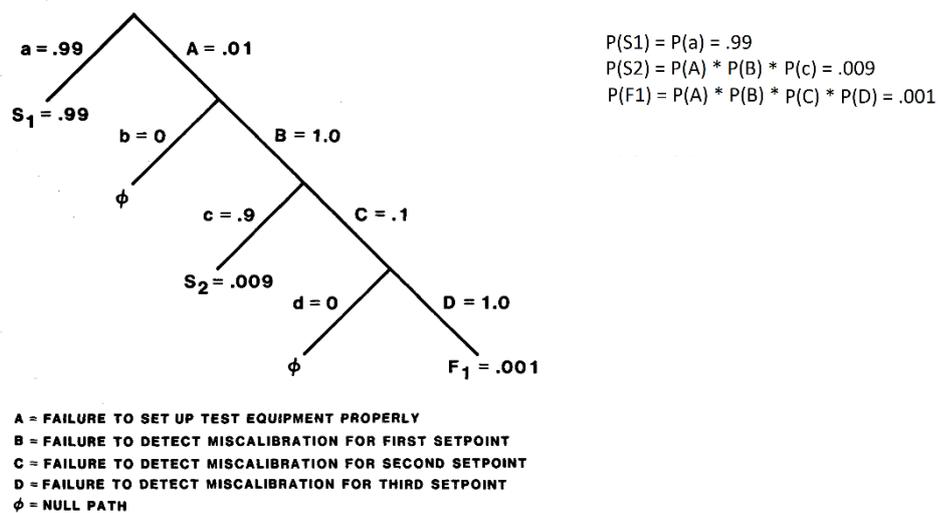

**Figure 2: Example of THERP binary event tree [33]**

THERP relies on a large human reliability database containing HEPs, which is based upon both plant data and expert judgments [33]. The nominal probability estimates from the analysis of the HRA event tree are modified for the effects of sequence-specific PSFs [18].

THERP models Errors of Omission (EOOs) and Errors of Commission (EOCs). Regarding cognitive error modeling, however, THERP uses available time to determine the probabilities of diagnosis failure and does not offer further breakdown in terms of specific cognitive or decision errors. A lack of detailed modeling of cognitive errors is common of first generation HRA methods and became the primary interest of the second-generation HRA methods [13].

THERP divides the PSF into three categories [13]: External PSFs, Internal PSFs, and Stressors. Although THERP provides a list of PIFs, it does not present specific rules for cause identification. In addition, it does not provide a specific procedure for identification of error mode [34]. The quantitative procedure of THERP is based on tables. Only three of the PSFs are used in HEP calculation: Tagging levels (of components or controls), experience, and stress.

*4.2 CREAM (Cognitive Reliability and Error Analysis Method)*

CREAM was developed by Hollnagel in 1993 [18]. It is a second-generation HRA method, and therefore focuses also on the cognitive aspect of human activity and its effect on performance. The basic principle of the model is a description of competence and control as two separate aspects of the performance. The distinction between competence and control is based upon Hollnagel's COCOM (contextual control) model [36], which defines:

- Competence: a person's skills and knowledge (the human capability to perform the task)
- Control: how the task is performed with the competence of the human. It is viewed as running along a continuum from a position where the individual has little/no control to where they have complete control. Several aspects of the context are identified; these are called Common Performance Conditions (CPCs).

The competence and control of an operator will determine the reliability of the performance [37]. Competence can be divided into observation, interpretation, planning and execution. Control, in its turn, can be divided into four control models:

- *Scrambled control*: when the task demand is high, the situation is unfamiliar and can change rapidly.
- *Opportunistic control*. The operator does no planning, which could be because the context is not clear or because of time constrains.
- *Tactical control*. The performance is based on a procedure and is planned.
- *Strategic control*. Operator looks further, with a wider horizon and considers the global context.

CREAM provides detailed instructions for both predictive and retrospective analyses. For the predictive task analyses, CREAM identifies several "basic human activities" into which a task can be decomposed. Each of these basic human activities corresponds to error modes. CREAM provides a list of nine PIFs (or CPCs) and their expected effect on performance reliability.

CREAM provides a two-level approach to calculate HEPs: a basic method and an extended method. The basic method steps are as follows:

1. Construction of event sequence and task analysis – analysis of detailed information of accident scenarios and the tasks to be conducted;
2. Assessment of the PIFs – or CPCs – each CPC is evaluated, and the combined score is calculated;
3. Determination of probable control mode – using **Error! Reference source not found.**.

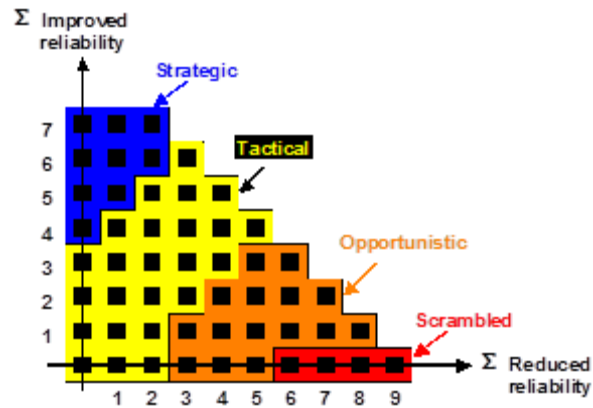

**Figure 3: CREAM control modes** [34]

For these four types of controls, the HEPs vary as follows:

$5E\text{-}6 < HEP(Strategic) < 1E\text{-}2$
$1E\text{-}3 < HEP(Tactical) < 1E\text{-}1$
$1E\text{-}2 < HEP(Opportunistic) < 5E\text{-}1$
$1E\text{-}1 < HEP(Scrambled) < 1$

The extended quantification method, for more detailed HEP assessments, can be seen in the method's manual.

### 4.3 HEART (Human Error Assessment and Reduction Technique)

HEART was first outlined by Williams in 1985, while working for the Central Electricity Generating Board [35]. The method was described in further detail in subsequent papers. HEART was designed to be a quick and simple method for quantifying the risk of human error – the method is designed to be used for HEP calculation only. HEART describes nine Generic Task Types (GTTs), each with an associated nominal human error potential (HEP). It describes each generic task with few sentences that specify the nature of the human action and its context. The analyst needs to identify the closest generic task to the task being analyzed [34]. HEART also presents 38 Error Producing Conditions (EPCs) that may affect task reliability, each has a maximum value by which the nominal HEP can be multiplied.

The key elements of the HEART method are:

1. The classification of the task for analysis into one of the nine Generic Task Types and assigning the nominal HEP to the task. The basic HEP values apply to these generic tasks are performed in "perfect" conditions. The HEPs are adjusted by the steps that follows when the generic tasks are performed in less than perfect conditions;
2. The identification of which EPCs may affect task reliability;
3. The assessment of the state of the EPCs by assigning a value ranged between 0 (best, positive) to 1 (worst, negative);
4. The calculation of the final HEP using Equation 1, in which "Effect" of the EPC corresponds to its weight factor

$$Final\ HEP = Basic\ HEP \times \prod_{i=1}^{n} \left[(Effect_{EPC\_i} - 1) \times State_{EFC\_i} + 1\right]$$

Equation 1

### *4.4 ATHEANA (A Technique for Human Event Analysis)*

ATHEANA is a second-generation method, and is the product of a multi-phase research sponsored by the US Nuclear Regulatory Commission [38]. The initial effort started in 1992 and aimed for more comprehensive coverage of operator response in the PRAs of nuclear power plants, particularly EOCs.

ATHEANA defines "error-forcing contexts" (EFCs), which are combinations of plant conditions and other influences that make an operator error more likely. An EFC is represented by a combination state of several PIFs. ATHEANA identifies the following PIFs to guide the experts to identify the EFCs:

- Procedures
- Training
- Communication
- Supervision
- Staffing
- Human-system interface
- Organizational factors
- Stress
- Environmental conditions
- Strategic factors such as multiple conflicting goals, time pressure, limited resources

The method can be used for both retrospective and prospective analyses. There are 10 steps in the ATHEANA HRA process, that can be summarized as [11]:

- The integration of the issues of concern into the ATHEANA HRA/PRA perspective;
- The identification of HFEs and unsafe actions that are relevant to the issue of concern;
- For each human failure event or unsafe action, the identification of the reasons why such events occur (plant conditions and performance shaping factors);
- The quantification of the EFCs and the probability of each unsafe action, given the context;
- The evaluation of the results of the analysis in terms of the issue for which the analysis was performed.

ATHEANA uses a quantification model for the probability of human failure events (HFEs) based upon estimates of how likely or frequently the plant conditions and PSFs comprising the EFCs occur. Subsequently, an expert elicitation approach for performing ATHEANA quantification was developed [39] because "… significant judgement must be exercised by the analysts performing the quantification. In fact, a significant amount of creativity and insight on the part of the analyst would be necessary to use existing HRA quantification methods to address the error-forcing conditions identified using ATHEANA". ATHEANA qualitative analysis is resource intensive and requires high levels of expertise and the assessment of HEP values mainly relies on expert judgment.

*4.5 SPAR-H (Standardized Plant Analysis Risk-Human Reliability Analysis)*

SPAR-H was first developed in 1994 by the U.S. Nuclear Regulatory Commission (NRC) in conjunction with the Idaho National Laboratory (INL). It was initially called Accident Sequence Precursor Standardized Plant Analysis Risk Model (ASP/SPAR). In 1999, based on the experience gained in field testing, this method was updated and renamed to its current denomination. The complete and current version was published in 2005 by the U.S.NRC [9].

The basic SPAR-H framework is the following [9]:

- It decomposes probability into contributions from diagnosis failures and action failures;
- It then accounts for the context associated with human failure events (HFEs) by using performance shaping factors (PSFs), and dependency assignment to adjust a base-case HEP;
- It uses pre-defined base-case HEPs and PSFs, together with guidance on how to assign the appropriate value of the PSF;
- It employs a beta distribution for uncertainty analysis, and
- Finally, it uses designated worksheets to ensure analyst consistency

The SPAR-H method assigns human activity to one of two general task categories: action or diagnosis. Eight PSFs are accounted for in the SPAR-H quantification process:

- Available time;
- Stress/Stressors;
- Complexity;
- Experience/Training;
- Procedures;
- Ergonomics/Human machine interface;
- Fitness for duty;
- Work processes

A major component of the SPAR H method is the SPAR H worksheet, which simplifies the estimation procedure. HEPs are determined by multiplicative calculation (i.e. Probability task failure x PSF1 X PSF2 x PSF3).

*4.6 Petro-HRA*

The Petro-HRA method has been developed as a result of an R&D project called "Analysis of Human Actions as Barriers in Major Accidents in the Petroleum Industry, Applicability of Human Reliability Analysis Methods" [40]. It is a method for qualitative and quantitative assessment of human reliability in the Oil and Gas industry, and it was built using SPAR-H as a basis.

Petro-HRA was developed in order to be integrated to a QRA, and consists of the steps illustrated in Figure 4.

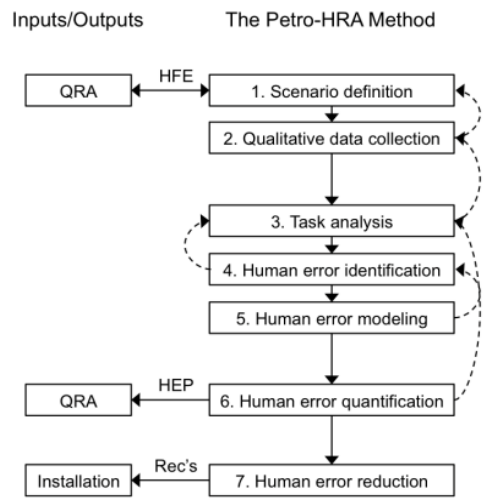

**Figure 4: Petro-HRA steps** [8]

The steps can be summarized as follows:

1. Scenario definition - In this step the analyst should attend or arrange several initial meetings. The Petro-HRA guideline contains several questions that can help the analyst in the meetings, as well as a list of documents that should be used to gather additional information.
2. Qualitative Data Collection - Collect specific and focused data from site visits, interviews and discussions with operators and documentation reviews, to enable a detailed task description.
3. Task Analysis - Describe the steps (i.e. human actions) that are carried out as part of an activity
4. Human error identification - Identify potential errors associated with task steps in the scenario, describe the likely consequences of each error, identify recovery opportunities, and describe the performance shaping factors (PSFs) that may have an impact on error probability.
5. Human error modelling. - Model the tasks in such a way that when individual tasks are quantified according to Step 6, the model logic can be used to calculate the HEP for the HFE that is then input to the QRA.
6. Human error quantification. Quantify each chosen task or event based on a nominal value and a set of PSFs. Check the reasonableness of the HEPs.
7. Human error reduction. Develop risk-informed improvement initiatives to reduce the human contribution to risk. Such improvements aim at either preventing the occurrence of human errors or mitigating their consequences.

The method recommends the use of the SHERPA taxonomy for error identification, presented in Table 1. It states, however, that the analyst can select another error taxonomy if this one better aligns with the types of tasks being modelled through the task analysis. Moreover, since the SHERPA taxonomy considers mainly action, checking, and communication errors, PETRO-HRA provides additional decision items that can be added to the SHERPA taxonomy (Table 2).

Table 1: The SHERPA error taxonomy [8]

| Action Errors | Checking Errors |
|---|---|
| A1-Operation too long/short | C1-Check omitted |
| A2-Operation mistimed | C2-Check incomplete |
| A3-Operation in wrong direction | C3-Right check on wrong object |
| A4-Operation too little/much | C4-Wrong check on right object |
| A5-Misalign | C5-Check mistimed |
| A6-Right operation on wrong object | C6-Wrong check on wrong object |
| A7-Wrong operation on right object | **Retrieval Errors** |
| A8-Operation omitted | R1-Information not obtained |
| A9-Operation incomplete | R2-Wrong information obtained |
| A10-Wrong operation on wrong object | R3-Information retrieval incomplete |
| **Information Communication Errors** | **Selection Errors** |
| I1-Information not communicated | S1-Selection omitted |
| I2-Wrong information communicated | S2-Wrong selection made |
| I3-Information communication incomplete | |

Table 2: Additional decision error taxonomy [8]

| Decision Errors |
|---|
| D1-Correct decision based on wrong/ missing information |
| D2-Incorrect decision based on right information |
| D3-Incorrect decision based on wrong/ missing information |
| D4-Failure to make a decision (impasse) |

For quantification, the PETRO-HRA uses a nominal human error probability (NHEP), as SPAR-H. The NHEP in Petro-HRA is 0.01 for all tasks, which is the same as for the diagnosis NHEP in SPAR-H. The separation between diagnosis (cognition) and action tasks in SPAR-H is not included in the Petro-HRA method because it was considered that all tasks are a combination of diagnosis and action.

The definitions of the PSFs in Petro-HRA differ from the ones of SPAR-H. Petro-HRA also modified the PSFs multipliers, and the guideline provides a discussion on these changes. The quantification follows the multiplication rule from SPAR-H. The guideline also provides a worksheet to aid the analyst in the quantification.

### *4.7 Phoenix-PRO (Phoenix for Petroleum Refining Operations)*

Phoenix is a model-based methodology developed by Ekanem in 2013 [41], who improved several aspects of previous works from different authors towards the development of a model based HRA methodology [25–28]. Phoenix was applied in oil refining scenarios in published papers [42–44]. Phoenix-PRO (Phoenix for Petroleum Refining Operations), in its turn, was developed as an adaptation of Phoenix for oil refining operations [7] arising from the need for an HRA method that reflects the particularities of this industry, and the ability of Phoenix to include these.

Phoenix analysis framework has three main layers: the "crew response tree" (CRT) (top layer), the human performance model (mid layer) – which uses fault trees, and the PIFs (bottom layer), as can be seen in Figure 5.

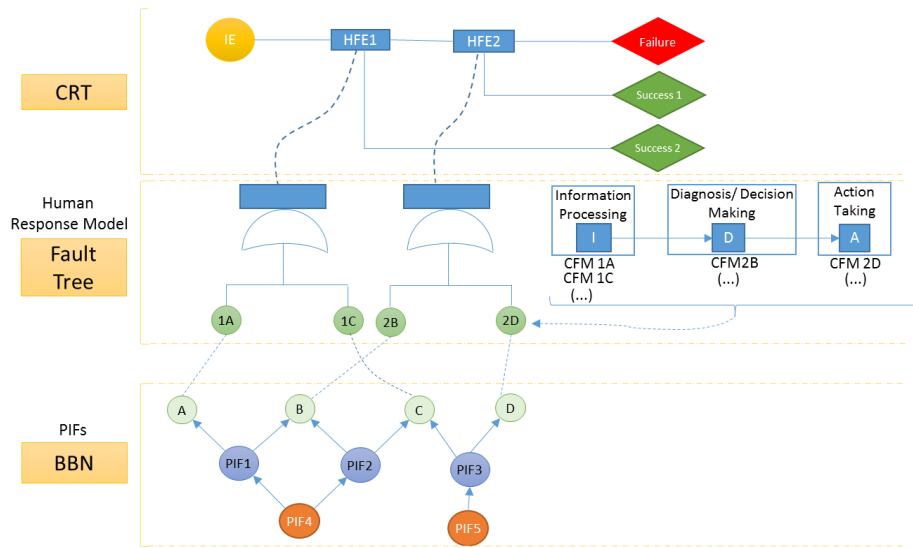

**Figure 5: Phoenix's layers [41]**

The same 3-layers structure is used in Phoenix-PRO. The three elements of the methodology are described below.

The crew response tree (CRT) is the first modeling tool for the qualitative analysis process. It is a forward branching tree of crew cognitive activities and actions, and acts as a crew-centric visual representation of the crew-plant scenarios. It provides the roadmap and blueprint that supports the performance and documentation of the qualitative analysis [10]. In the CRT, each sequence of events indicates one of the possible crew responses across the entire accident sequence. This helps increasing consistency and reducing variability in the HRA task analysis [41].The CRTs are developed to model HFEs corresponding to a given safety function.

Clearly, in order to identify the safety function and construct the CRT the analyst must be familiarized with the scenario, with aid of documentation about the process and the crew and all procedures used to carry out the safety function. Phoenix-PRO provides a guidance list of needed information and documentation for the analyst [45]. Moreover, Phoenix-PRO provides a flowchart in order to aid the analyst to construct the CRTs, with questions to guide the addition of branches to the CRT.

The mid layer comprehends the human response model – the human failure mechanisms and their causes. It captures the remaining aspects of the context through a set of supporting models of crew behavior in the form of causal trees (fault trees) that are linked to the CRT branches. Phoenix uses the crew centered version of the Information, Decision and Action (IDA) [46] cognitive model as the basis for this linkage. IDA is a three-stage model and these stages serve as the basis for linking failure mechanisms to the possible human failures. The IDA stages are as follows [46]:

- I - Information pre-processing: This phase refers to the highly automatic process of processing incoming information. It includes information filtering, comprehension and retrieval;
- D - Diagnosis/ Decision making: In this phase the crew uses the perceived information and the cues from the previous stage, along with stored memories, knowledge and experience to understand and develop a mental model of the situation. In addition, the crew engages in decision making strategies to plan the appropriate course of action.
- A - Action: In this final phase the crew executes the decision made through the D process

The crew errors in Phoenix are defined through the IDA stages. This means that the crew can:

- fail while information gathering (I stage);
- have the correct and complete information in hand, but fail in situation assessment, problem solving and decision making (D stage);
- have the correct and complete information and make a correct decision but fail to execute the correct action (A stage).

Thus, Crew Failures Modes (CFMs) are used to further specify the possible forms of failure in each of the Information, Decision and Action phases (i.e. they represent the manner in which failures occur in each IDA phase). The CFMs used in Phoenix-PRO are presented in Table 3.

When an analysis is conducted in the context of a particular scenario, depending on the I-D-A phase, only a subset of the CFMs will apply. Phoenix provides an initial set of fault trees to aid analysts in selecting the relevant CFMs for each branch point within each scenario [41].

**Table 3: Phoenix-PRO Crew Failure Modes**

| ID | Crew Failure Modes in "I" Phase | ID | Crew Failure Modes in "D" Phase | ID | Crew Failure Modes in "A" Phase |
|---|---|---|---|---|---|
| I1 | Key Alarm / Information not Responded to (intentional & unintentional) | D1 | Plant/System State Misdiagnosed | A1 | Incorrect Timing of Action |
| I2 | Data Not Obtained (Intentional) | D2 | Procedure Misinterpreted | A2 | Incorrect Operation on Component/Object |
| I3 | Data Discounted | D3 | Failure to Adapt Procedures to the situation | A3 | Action on Wrong Component / Object |
| I4 | Data Incorrectly Processed | D4 | Procedure Step Omitted (Intentional) | | |
| I5 | Reading Error | D5 | Procedure not followed | | |
| I6 | Information Miscommunicated | D6 | Inappropriate Procedure Followed | | |
| I7 | Wrong Data Source Attended to | D7 | Decision to Delay Action | | |
| I8 | Data Not Checked with Appropriate Frequency | D8 | Inappropriate Strategy Chosen | | |

The third layer of the methodology comprises the PIFs – presented in Table 4. The CFMs and the PIFs are connected through Bayesian Belief Networks (BBNs). The BBN models paths of influence of the PIFs on each other and on the various CFMs. The PIFs have been organized into nine (8) main groups, which are also individually considered as PIFs - "Primary or level 1

PIFs". The PIFs are classified into levels within the groups, thus forming a hierarchical structure, which can be fully expanded for use in qualitative analysis as well as collapsed for use in quantitative analysis. In this hierarchy, Level 1 PIFs affect directly the CFMs. Level 2 PIFs affect level 1 PIFs, and level 3 PIFs affect Level 2.

Phoenix's qualitative analysis main steps are:

1. Scenario Development / Familiarization
2. Development of Crew Response Tree – The development of the CRT leverages from a flowchart to ensure traceability.
3. Identification of Crew Failure Modes – For each HFE identified in the CRT, the analyst identified the relevant CFMs using the FTs.
4. Identification of relevant PIFs for the CFMs – For each CFM, the analyst identifies the applicable PIFs.
5. Model integration and analysis of HFE Scenarios, Development of Narratives, and Identification of Dependencies

The qualitative analysis provides the CFM cut-sets and the relevant PIFs as inputs for the quantitative analysis. The quantitative analysis process comprehends the following steps:

1. Assess and estimate relevant PIFs levels – Phoenix provides a set of questionnaires for assessing the state of the relevant PIFs;
2. Input the estimated levels into the model;
3. Determine if dependency is to be considered;
4. Follow procedure for dependency or non-dependency quantification;
5. Estimate the conditional probabilities of the CFMs – this step is done using the BBNs;
6. Estimate the HEP for the specific HFE. During this step the probabilities of the HFEs of the CRT are calculated, given the relevant CFMs and using the FTs logic;
7. Estimate the probability of the end states of the CRT, given the HFEs.

The linking of the CRT, FTs and BBNs, i.e. the interaction between crew and technical system, human response model and causal model, allows then for assessing the contribution of CMFs or PIFs to the final risk.

**Table 4: Phoenix-PRO Performance Influencing Factors**

| HMI | Procedures | Resources | Team Effectiveness | Knowledge / Abilities | Bias | Stress | Task Load |
|---|---|---|---|---|---|---|---|
| HMI Input | Procedure Content | Tools | Communication | Knowledge / Experience / Skill (content) | Motivation/ Commitment | Stress due to Situation Perception | Cognitive Complexity |
| HMI Output | Procedure Updating | Tool Availability | Communication Quality | Task Training | Confidence in Instruments | Perceived Situation Urgency | Inherent Cognitive Complexity |
| | Procedure Availability | Tool Quality | Communication Availability | Knowledge of Plant Conditions | Familiarity with of Recency of Situation | Perceived Situation Severity | Cognitive Complexity due to External Factors |

|  |  | Workplace Adequacy | Team Coordination | Knowledge / Experience / Skill (access) | Competing or Conflicting Goals | Stress due to Decision | Execution Complexity |
|---|---|---|---|---|---|---|---|
|  |  |  | Leadership | Attention |  |  | Inherent Execution Complexity |
|  |  |  | Team Cohesion | Fitness for Duty |  |  | Execution Complexity due to External Factors |
|  |  |  | Responsibility Awareness |  |  |  | Extra Work Load |
|  |  |  | Team Composition |  |  |  | Passive Information Load |
|  |  |  | Team Training |  |  |  |  |

| Key | Meaning |
|---|---|
|  | Level 1 PIFS |
|  | Level 2 PIFS |
|  | Level 3PIFS |

The application of Phoenix can leverage from the use of a software for modeling and quantification, such as the Integrated Risk Information System (IRIS)[**]. Current efforts at the B. John Garrick Institute for the Risk Sciences at UCLA include the development of a dedicated software for Phoenix.

## 5 HRA for Oil and Gas Operations: Methods Evaluation

This section presents an evaluation of the methods described in Section 4 for their applicability and potential use as a foundation for a method for Oil and Gas industry. The evaluation was performed by the authors, considering their experience on HRA and on oil and gas operations. This evaluation was performed in three stages. First, the methods were evaluated regarding their core characteristics and suitability for the Oil and Gas industry aiming at assessing how well they could represent or be adapted for the industry. Second, the methods were evaluated against desirable features of an HRA method, deriving from HRA best practices. Third, the high-ranked methods of the previous stage were evaluated regarding how well they comply with key attributes of a robust state-of-the-art methodology. i.e., how they can be accurate, reproducible and traceable regarding its analysis of operator performance, and, consequently, how efficient they can be as a basis for risk-informed decisions. These attributes were derived from the discussions presented in Sections 3.2 and 3.3.

It should be noted that the Oil and Gas industry comprises operations that can highly differ among themselves, such as control room operations, field operations, emergency response, onshore and offshore settings, and upstream, midstream and downstream specific activities.

---

[**] More information at https://www.risksciences.ucla.edu/software/iris

This review does not differentiate these operations at this point and considers the general characteristics they share.

### 5.1 First Stage: HRA methods evaluation for suitability for oil and gas operations

Regarding the potential for application in Oil and Gas, the methods are assessed in terms of degree of required adaptations. This relates to estimated time consumption for the adaptations and capability of the method to absorb it in its framework. In this sense, a low degree of required adaptations means a high suitability of the method. Table 5 presents the final assessment as low, medium and high. Note that this is a qualitative assessment. Some methods marked as "low suitability" have been successfully used before in HRAs of oil and gas operations, and this assessment does not aim at questioning the validity of these applications. Rather than an application of the method in one specific scenario and context, the assessment aims at evaluating the methods as a foundation for an HRA method tailored for oil and gas operations.

**Table 5: Suitability of methods for Oil and Gas applications**

| Method | Suitability in terms of required adaptations | Comments |
|---|---|---|
| **THERP** | Low | THERP uses "available time" to determine the probabilities of diagnosis failure, and no further breakdown in terms of specific cognitive or decision errors is offered. However, the available time windows for action in nuclear power plant operation and petroleum operations are often significantly different. Recovery time windows for nuclear accidents may vary from hours to days [34]. In oil refineries, in contrast, the time windows for action may be very short—e.g. time for action after the leak of a flammable gas. This difference cast doubt on the applicability of the HEP estimates based mostly on nuclear reactor time scales.<br><br>In addition, THERP has a fixed search scheme to identify possible errors and calculate HEPs for activities in nuclear power plants. It would require significant revision in terms of search scheme, parameter values, and adding new performance shaping factors (PSFs) for use in petroleum operations. It does not offer enough guidance on modelling scenarios and the impact of PSFs on performance [47]. |
| **HEART** | Medium | HEART uses only nine Generic Tasks, which may not be enough to cover all petroleum operations human activities. Moreover, although the nonspecific task classification scheme makes HEART applicable to a wide range of industries, it can also make it difficult for analysts to identify a petroleum industry operation task as one of the generic tasks. A variation of HEART specific for NPP operations, NARA, uses different weighs for some of the error producing conditions. This suggests that the HEART weights and perhaps the HEPs of the general tasks would need to be revisited for petroleum applications |

| Method | Suitability in terms of required adaptations | Comments |
|---|---|---|
| **CREAM** | High | Although CREAM was developed for use in the nuclear industry, the underlying method is generic and, therefore, it could be applied in petroleum operation. This would require, however, an analysis of its nine PSFs in order to evaluate if they cover all factors that can affect human performance in this industry, in addition to a revision on the probabilities used by the model to analyze its adequacy to petroleum industry operations. |
| **ATHEANA** | Medium | ATHEANA is an expert judgment based HRA method. Since the ATHEANA guidelines for searching for error forcing contexts were developed for nuclear operations, new guidelines would need to be developed for petroleum industry operation tasks. US NRC report NUREG-162428 states that the effectiveness of the ATHEANA methodology results from forming a diverse, experienced project team to perform a comprehensive, broad-ranging analysis. Few organizations are able to undertake such an extensive analysis without clearly defined, commensurate benefits. ATHEANA could be suitable for petroleum industry operations but would require considerable effort for adaptation. |
| **SPAR-H** | High | Since SPAR-H worksheets are designed for nuclear power operations, the worksheets need to be revised regarding the scope of PIFs and their corresponding weights. The Petro-HRA method is based on SPAR-H. |
| **Petro-HRA** | High | Petro-HRA methodology was developed for the oil and gas industry. It is mainly intended for use within a QRA framework but may also be used as a stand- alone analysis. It is not validated yet for use in pre-initiating scenarios but can be used to estimate the likelihood of human failure events (HFEs) in post -initiating event scenarios. Petro-HRA is a suitable method for use in analysis of the petroleum industry operation, since it is already endorsed by a major Oil company – Statoil and provides solid documentation. |
| **Phoenix-PRO** | High | Although Phoenix was originally developed using elements related to Nuclear Power Plants operations, it was object of adaptations for use in oil refineries operation resulting in the development of Phoenix-PRO. The development of Phoenix-PRO was based, especially, but not limited to, on studies of past accidents in oil refineries, visitations to the integrated control room of an oil refinery, and experts' opinions and feedback obtained through a questionnaire. The main changes in this adaptation were in the CRT flowchart, CFMs and the PIFs definitions. The adaptations focused on oil refineries and petrochemical plants operations. They need to be reviewed in order to include particularities of upstream and midstream operations. They can also benefit of review using more evidences – coming from incident reports, interviews, and plants visitations. |

## 5.2 *Second Stage: HRA methods evaluation against good practices and desirable attributes*

The second stage consists of an initial screening, in which we elected five items that are desirable in an HRA method and evaluated how these items are presented in the HRA methods - as high, medium or low. The items are derived from the good practices stated at NUREG-1792.

The methods' evaluated items are as follows:

A. Task analysis and decomposition - *Task decomposition is a key process to break down the human activity to match the method's unit of analysis for human error*
B. PIFs identification and guidance on how to assess them and how they affect the Human Error Probability
C. Assessment of dependency
D. Human Error Identification - *Providing procedure for human error identification ensures that different analysts applying the method for the same scenario will consider the same human errors*
E. Reproducibility of HEP estimation - *Given the method's procedure for HEP estimation, how reproducible the results are when applied by different analysts?*
F. Potential for adaptations for application in Oil and Gas operations (see comments in Table 5)

Table 6 summarizes the assessment of the methods regarding the desirable items for an HRA method. Note that it is possible that a method does present an attribute, e.g. consideration of dependency, but does not do it in a desirable level. THERP, for example, allows consideration of dependency, but does not provide clear rules for the assessment of the level of dependency. This makes the assessment a rather difficult task, requiring a considerable amount of expert judgment, which may lack transparency and traceability and leads to low repeatability of the results. For this attribute, then, THERP scores "medium".

The overall assessment of the methods is indicated by the sum of each attribute, scoring 0 to "low", 1 to "medium" and 2 to "high".

**Table 6: Evaluation of HRA methods items against desirable attributes**

| Method | Items | | | | | | |
|---|---|---|---|---|---|---|---|
| | A | B | C | D | E | F | Score |
| **THERP** | High<br><br>*Decomposes tasks into Screening, Diagnosis and Action* | Low<br><br>*THERP provides a list of PSFs but gives no specific rules to assess the states of these PSFs and their effects on HEPs.* | Medium<br><br>*THERP provides five levels of dependency between two consecutive operator activities* | Low<br><br>*Does not provide explicit procedures for performing error identification* | Low<br><br>*The use of a simple, generic TRC for addressing diagnosis errors is an over-simplification for addressing cognitive causes and failure rates for diagnosis errors* | Low | 3 |
| **HEART** | Medium<br><br>*HEART does not provide an explicit procedure for task decomposition. But it specifies nine generic tasks for the analyst to identify the best-matched generic task for the task of interest.* | Medium<br><br>*HEART provides a long list of PSFs that can be used to modify the basic HEPs. No causal model is provided for the identification of "root causes"* | Medium<br><br>*The effects of task dependencies are implicitly embedded in the definitions of the generic tasks* | Low<br><br>*No explicit procedures for error identification* | Low<br><br>*There is the possibility that different analysts will select a different generic task (and therefore different HEPs) for the same task* | Medium | 4 |
| **CREAM** | High<br><br>*CREAM identifies fifteen basic tasks to decompose the human activities of interest* | High<br><br>*CREAM identifies a list of nine Common Performance Conditions (CPCs) (similar to PSFs) that could affect HEPs* | Low<br><br>*CREAM does not provide a specific procedure for identifying and accounting for task or error dependencies* | Low<br><br>*No explicit procedures for error identification* | High<br><br>*Decomposing the analysis into a limited set of subtasks defined by the basic human activities is relatively straightforward, and reproducibility is high.* | High | 8 |

| | | | | | | | | |
|---|---|---|---|---|---|---|---|---|
| **ATHEANA** | Low<br><br>*Not specified* | Low<br><br>*User-defined* | Medium<br><br>*The flexibility of the framework allows experts to define the scenarios in terms of possibly interdependent tasks and consider the impact on HEP assessment.* | Low<br><br>*ATHEANA has many ambiguous steps that make it hard to follow. The result means that reproducibility of ATHEANA is highly expert-dependent. reached* | Low<br><br>*HEP quantification in ATHEANA is expert opinion based. May be difficult to trace or reproduce the origins of experts' HEP estimates* | Medium | 2 |
| **SPAR-H** | High<br><br>*SPAR-H decomposes a task into subtask of "diagnosis" and/or "action."* | Medium<br><br>*SPAR-H provides a set of PIFs and how to assess them. Resolution of the PSFs may be inadequate for detailed analysis* | High<br><br>*SPAR-H provides guidelines to assess the level of dependency of actions.* | Low<br><br>*SPAR-H does not provide guidelines for error identification* | Medium<br><br>*Although authors checked underlying data for consistency with other methods, basis for selection of final values was not always clear. The method may not be appropriate where more realistic, detailed analysis of diagnosis errors is needed.* | Medium | 7 |
| **Petro-HRA** | High<br><br>*Petro-HRA provides a guideline for task decomposition* | High<br><br>*Petro-HRA uses a modified version of the SPAR-H set* | High<br><br>*Petro-HRA uses similar framework for dependency as SPAR-H* | High<br><br>*Petro-HRA provides guideline for error identification using the SHERPA taxonomy with additional decision errors* | Medium<br><br>*Petro-HRA modified SPAR-H multipliers and provides a straightforward method for quantification of HEP. Without a causal model, though, it is possible that different analysts chose different PIFs and have a different final HEP* | High | 11 |

| Phoenix-PRO | High | High | High | High | High | High | 12 |
|---|---|---|---|---|---|---|---|
| | *Phoenix provides task decomposition using the cognitive model IDAC* | *Phoenix-PRO provides a model-based PIF list and calculates its effect on HEP through Bayesian Belief Networks* | *Phoenix-PRO provides a model-based framework to consider dependency, using dynamic BBN. This needs to be further validated.* | *Phoenix-PRO provides a model-based guideline to identify human errors through the Crew Response Tree* | *HEP quantification is done using a Hybrid Casual Logic modeling, through Event Sequence Diagrams, Fault Trees and BBNs.* | | |

## 5.3 Third Stage: HRA methods evaluation against requirements for advanced HRA methods

The methods that presented a higher ranking in the first two stages are the ones that i) present HRA desirable items in an adequate level ; ii) have a high potential for applications on oil and gas operations. These attributes, however, dot not assure the robustness of the method, particularly concerning traceability and reproducibility. Therefore, the methods with the higher rankings from Stages 1 and 2 were assessed against the attributes that are necessary for advanced HRA methods, in Stage 3. These attributes were further discussed in Sections 3.2 and 3.3.

A robust HRA methodology is one that provides, among other features, reproducibility and traceability. These features can ensure that, when applied to the same scenario by different analysts, the methodology will not result in different qualitative or quantitative conclusions. Section 3 provided the attributes of a robust HRA methodology, and pointed that, in brief: i) a robust methodology should have a causal model, to provide both the explanatory and predictive capabilities; and ii) a robust methodology should be model-based approach, to ensure reproducibility of the results, and robustness of the predictions.

The three methods with a higher ranking on Stage 1 of this assessment – CREAM, Petro-HRA and Phoenix-PRO, were evaluated regarding three main attributes for a robust HRA methodology: a causal model that provides explanatory power, reliability and reproducibly, and traceability. Table 7 presents the results.

**Table 7: Assessment of high-ranking methods against the attributes of a robust HRA methodology**

| No | Attributes of an advanced HRA Methodology | Phoenix-PRO | Petro-HRA | CREAM |
|---|---|---|---|---|
| 1 | Explanatory power, "causal model" for error mechanisms and relation to context, theoretical foundations | Yes | No | Partial |
| 2 | Be informed by both data and models, and able to incorporate multiple sources. | Yes | No | No |
| 3 | Be flexible enough to accommodate changes in database structures, data sources. | Yes | Yes | No |
| 4 | Reliability (Reproducibly, Consistency, Inter- and Intra-rater Reliability) | Yes | No | No |
| 5 | Traceability/Transparency | Yes | Yes | No |

Phoenix-PRO, as seen in Table 7, presents the main attributes of a robust HRA methodology. Indeed, Phoenix was developed using as a foundation a model-based HRA framework [25,26,28] that was introduced as a response for the requirements for an advanced HRA methodology.

Phoenix was, thus, developed as a response to the need of a methodology that would overcome the limitations of the previous ones, and acknowledging the best practices pointed by NUREG-1792 and other authors. It assimilates strong elements of current HRA good practices and adopts lessons learned from empirical studies. Phoenix-PRO adopts the structure and framework of Phoenix – and the conclusions drawn on Phoenix are applicable for Phoenix-PRO. Note that the assessment of Phoenix as a solid foundation for developing a method for Oil and Gas is not without limitations. These are discussed in the following Section.

## 6 Discussion and Concluding thoughts

The safety of Oil and Gas operations have been increasing with the adoption of advanced control loops, improvements in safety culture and organizational factors, use of reliable equipment and instruments and implementation of complex safety barriers. The industry, in general, applies advanced techniques for process safety, quantitative risk assessment, quantitative models for consequence evaluation such as clouds dispersion and fire propagation. Human error, however, has not yet been addressed in the industry in a systematic manner. The need of an HRA method specific for oil and gas operations is driven by the differences this industry presents in comparison to other industries, in particular to the Nuclear industry – main user and developer of HRA methods.

The need of performing HRA and reducing human error led to recent initiatives concerning the development of an HRA method tailored for Oil and Gas operations. These initiatives follow the principle of adapting an existing method considering the particularities of the industry. Choosing a method as a foundation for this development, however, should be based not only on the potentiality of the method to be used in the industry, but also on its features regarding advanced HRA methods.

The Nuclear industry has an accumulated experience with using HRA, and this experience should be leveraged by the Oil and Gas industry into choosing a method for foundation. In general, the current momentum for enhancing HRA benefits from an increasing understanding of human behavior and the need of including cognitive sciences in HRA, the HRA data collection initiatives, the development of robust quantification methods, and the possibilities of using software for complex analyses. Current discussions on third generation HRA methods leverage those points and remark that the method should be model-based, comprehensive, and flexible to accommodate simulator data, among other features.

The state-of-the-art discussions on HRA, led by the Nuclear industry community, must be seriously considered in defining a technical roadmap to a credible HRA method for the Oil and Gas industry. The assessment of the methods against potential adaptability to Oil and Gas operations, desirable features of an HRA method, and requirements for advanced methods pointed at Phoenix-PRO, an adaptation of Phoenix, as a potential candidate for foundation for a method tailored for Oil and Gas industry. It should be noted that this study does not aim at questioning the validity of specific uses of other methods in oil and gas operations, or other methods developed for the industry. Rather, its main objective is to point the need for acknowledging and accommodating state of the art discussion within the HRA community for HRA applications in the Oil and Gas industry.

Phoenix is a model-based method developed for responding to the requirements of an advanced HRA method. It is sufficiently flexible for accommodating adaptations to represent oil and gas operations. Moreover, it was designed for being able to accommodate data collected in current projects such as SACADA. It makes use of cognitive model for the operator, and uses a robust quantification approach, in addition to a traceable qualitative method. Current efforts on Phoenix include the enhancement of its causal model and the development of a dedicated software.

It should be noted that the conclusions of this study are not definitive. Despite providing a solid foundation for a dedicated HRA method for the industry and having been subject of application to some scenarios of the industry, Phoenix must still be assessed regarding its use within QRA. In addition, the method still lacks extended industry applications, due to the recency of its development. This is subject of current work being developed by some of the authors

25[36] Hollnagel E. Human reliability analysis: Context and control. Academic Press; 1993.

[37] Hogenboom I, Kristensen AS. Comparison of Human Reliability Analysis Method 2018.

[38] US Nuclear Regulatory Commission (USNRC). Technical Basis and Implementation Guidelines for A Technique for Human Event Analysis (ATHEANA). Washington: 2000.

[39] Forester J, Bley D, Cooper S, Lois E, Siu N, Kolaczkowski A, et al. Expert elicitation approach for performing ATHEANA quantification. Reliab Eng Syst Saf 2004;83:207–20. doi:10.1016/j.ress.2003.09.011.

[40] Bye A, Laumann K, Taylor C, Rasmussen M, Øie S, Merwe K van de, et al. The Petro-HRA Guideline. 2017.

[41] Ekanem N. A Model-Based Human Reliabiliyt Analysis Methodology (PHOENIX Method). University of Maryland, 2013.

[42] Ramos MA, Droguett EL, Mosleh A. Human Reliability Analysis of Past Oil Refinery Accidents using Phoenix Methodology. Proc. ABRISCO 2015-PSAM Top. Meet. Saf. Reliab. Oil Gas Explor. Prod. Act., 2012.

[43] Ramos MA, Droguett EL, Mosleh A, Moura M das C, Martins MR. Revisiting past refinery accidents from a human reliability analysis perspective: The BP Texas City and the Chevron Richmond accidents. Can J Chem Eng 2017;9999:2293–305. doi:10.1002/cjce.22996.

[44] Ramos MA, Droguett EL, Mosleh A, Martins MR. Revisiting Past Refinery Accidents From a Human Reliability Analysis Perspective. An. do XXI Congr. Bras. Eng. Química, Fortaleza: 2016.

[45] Ramos MA. A Methodology for Human Reliability Analysis of Oil Refineries and Petrochemical Plants. Federal University of Pernambuco, 2017.

[46] Chang YHJ, Mosleh A. Cognitive modeling and dynamic probabilistic simulation of operating crew response to complex system accidents Part 1 : Overview of the IDAC Model 2007;92:997–1013. doi:10.1016/j.ress.2006.05.014.

[47] Kirwan B. A guide to practical human reliability assessment. London: CRC Press; 1994.

[48] James Chang Y, Bley D, Criscione L, Kirwan B, Mosleh A, Madary T, et al. The SACADA database for human reliability and human performance. Reliab Eng Syst Saf 2014;125:117–33. doi:10.1016/j.ress.2013.07.014.